\documentclass[11pt,a4paper]{article}
\usepackage{geometry}
\usepackage{times}
\usepackage{latexsym}
\usepackage{xspace}
\usepackage{multirow}
\usepackage{url}
\usepackage{booktabs}
\usepackage{tikz,tikz-qtree}
\usepackage{pgfplots}
\usepackage{amssymb}
\usepackage{xfrac}
\usepackage{graphicx}
\usepackage{tablefootnote}
\usepackage{amsmath}
\usepackage{enumitem}
\usepackage{multicol}
\usepackage{hyperref}

\pgfplotsset{compat=1.14}
\usepackage{tabularx} 
\newcommand{\myfloatalign}{\centering} 
\usepackage{caption}
\usepackage{subfig}  

\usepackage{classicthesis}

\author{Shirui Tang \\
{\tt shirui.tang@cantab.net}}
\title{Dynamics of feed forward induced interference training}
\date{\today}

\begin{document}
\maketitle

\begin{abstract}
  \noindent Preceptron model updating with back propagation has become the routine of deep learning. Continuous feed forward procedure is required in order for backward propagate to function properly.
  Doubting the underlying physical interpretation on transformer based models such as GPT brought about by the routine explaination, a new method of training is proposed in order to keep self-consistency of the physics.
  By treating the GPT model as a space-time diagram, and then trace the worldlines of signals, identifing the possible paths of signals in order fot a self-attention event to occure. With a slight modification, self-attention can be viewed as an ising model interaction, 
  which enables the goal to be designed as energy of system. Target is treated as an external magnetic field inducing signals modeled as magnetic dipoles. A probability network is designed to pilot input signals travelling for different durations through different routes. 
  A rule of updating the probabilities is designed in order to form constructive interference at target locations so that instantaneous energy can be maximised. 
  Experiment is conducted on a 4-class classification problem extracted from MNIST. The results exhibit interesting but expected behavours, which do not exist in a bp updated network, but more like learning in a real human, especially in the few-shot scenario. 
\end{abstract}

\section{Introduction}

Self-attention is perhaps the most successful idea in nlp tasks. After realizing the physical process of self-attention is the positive feedback interaction, 
ising model becomes a viable model to describe a modified self-attention process. Inspired by the physical nature of self-attention, more physics restrictions 
were applied to convert the routine preceptron model classification task from a mathematical optimization problem into a physics problem involving magnetic force and optic interference. 

  Self-attention was regarded as a long-ranged global interaction process \cite{DBLP:journals/corr/abs-1804-09541}, with the measuring distance being spatial or temporal. However if the metric distance in vector phase space is being considered, 
then the interactions would take place between similar neighbours; the interaction strength would decreases with metric distance. Therefore the interaction can be treated as local in phase space.
Magnetic force in ising model is also a local interaction. Another common property between ising model and self-attention is positive feedback, 
letting two similar vectors attend to each other to be updated and become even more similar.

Softmax attention weights can be turned into 'hard' weights to increase interpretability of model \cite{xu2015show} \cite{cui2019fine}. 
Consider a sphere in phase space centered at query vector's point, and uniform attention probabilities are distributed to key vectors near the query within a threshold radius; any key outside the sphere are defined 'non-similar', and receives zero attention weight. 
Now self-attention is simplified to ising model \cite{ising}. 
Feature vectors are normalized to length 1, so that they can be treated as magnetic dipoles with dipole moment equals 1. 
These dipoles are called signals, each has an emision source with well-defined spatial coordinate. A source can be considered as a small patch of retina cells, emiting multiple signals at each timestamp, with phase space directions changing with time. 
If the signals are indeed interacting to each other like magnetic dipoles in ising model, then the phase-similar but spatially distant dipoles must be brought together in physical space. 
Physics rule restricts dipoles can only exert forces within a limited spatial range at any time. 
The gathering dipoles are the travelling signals; they may have different emision sources, all arriving a common destination, where attention interaction physically take place. This defines a self-attention interaction event in spacetime diagram. 
Clearly the system becomes dynamic because emision and interaction events will have different coordinates in the spacetime diagram. 

Put the GPT \cite{Radford2018ImprovingLU} model into a space-time diagram, 
with time on the x-axis. Any input query vector emits copies of signals, they are teleported to its output position instantaneously, tracing a vertical worldline. 
All the key vectors interacting with the query must trace worldlines intersecting at one same point on the query worldline to allow attention mechanism physically take place. 
Earlier signals can travel to their future but they can not travel backward in time to take part a self-attention interaction event held before their emision time. 
Every timestamp in every target position, there exists a self-attention event. Excluding the ones not possible to arrive in time, a signal needs to make a choice over which event should it join. 
Hence there is a probability associated with each worldline leading to a destination event. 

It is assumed that multiple signals can be emitted at one timestamp for each source, all of them are identical to their normalized input vector at input time. 
Signals travelling different worldlines according to a probability distribution. In a self-attention interaction event, signals from different space-time sources 
interact to each other, the similarity (dot product) between normalized signals is in the range $[-1, +1]$, and our goal is to maximize the total similarities among all signals in this self-attention event. 
Treating the self-attention event as a ising model interaction 'meeting', signals can interact to each other freely, or they can be influence by a target signal. In ising model target is the external magnetic field, interacting to every dipole signal. 
In the meeting analogy, target is the host of meeting event, selecting his prferable guests who are obviously more similar to the him. 
By telling the sources to increase or decrease the probability of sending signals (guests) to his meeting, a host can maximize the total similarity score, which is physically defined as the energy of ising model. 
Once the probabilities are learned, sources will send similar guests to the meeting even the host is absence; this can be described by a constructive interference process for the arriving signals.\cite{BornWolf:1999:Book} 
Our energy model considers only the host-guest interactions, but not the guest-guest interactions; this helps with saving computing expense. The theory involving guest-guest interaction will be more complicated as it will be discussed in the Further Work section.

~\\
\section{Background}

  \subsection{Optics}
\begin{figure*}
{\includegraphics[width=1.0\linewidth]{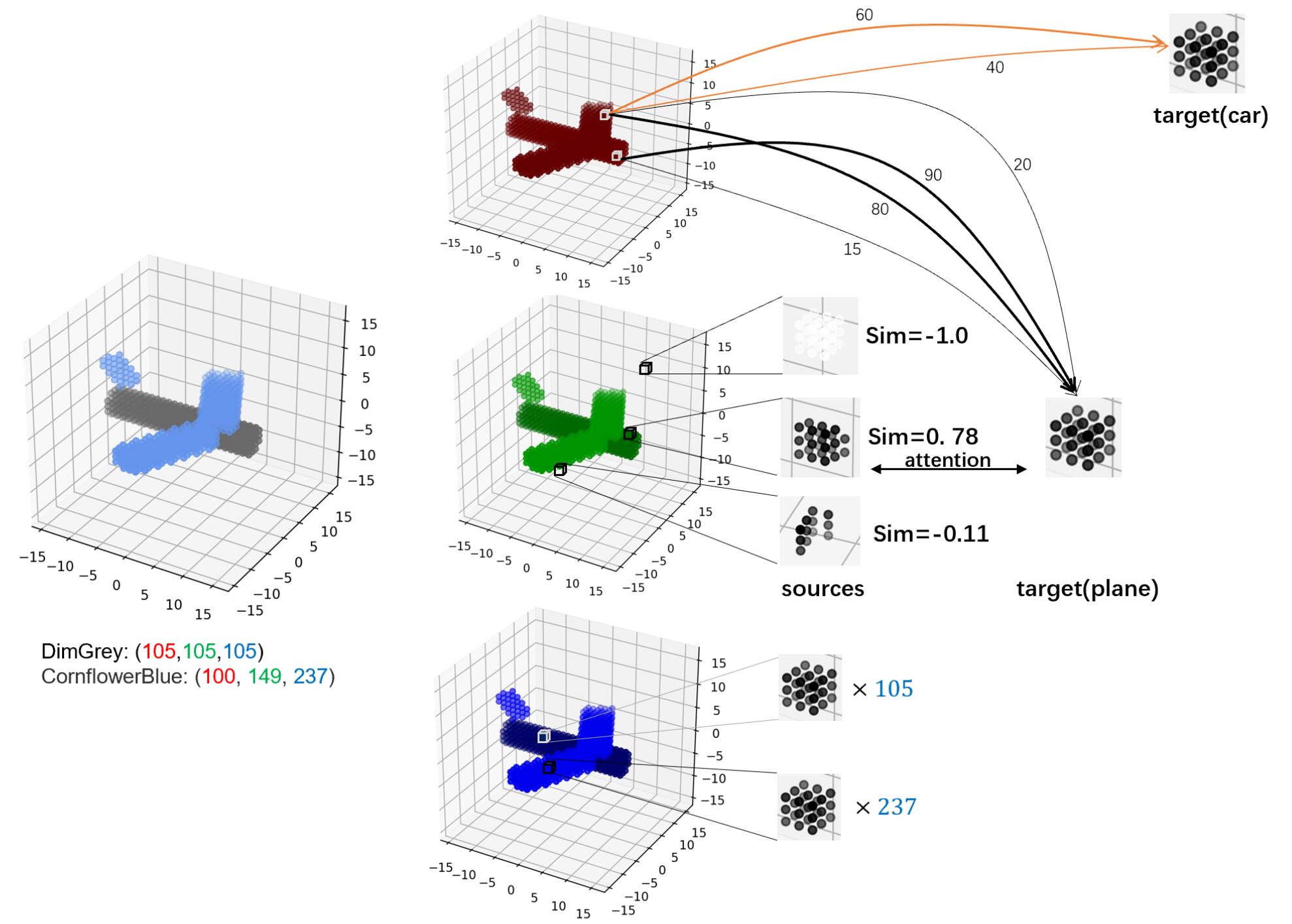}}
\caption[generation, conduction and interference of signals]{generation, conduction and interference of signals}\label{fig:setup}
\end{figure*}

  In CV the RGB are normally represented as three channels of features, mapping the input picture to one point in phase space. 
  Intensities of channels are linear transformed with a 3x3 kernel, meaning the magnitudes can be arbitrarily scaled up or down and then transmitted into the next neuron. 
  However in neuroscience, the action potential of neural transmition does not represent the magnitude; frequency of signal is more related to the magnitude of stimulation \cite{wiki:neural_coding}. 
  Therefore three postulates are made:
  \begin{enumerate}
    \item Each neuron has a binary value, +1 or -1, either on or off. Small fluctuations around the two values are allowed. 
    \item The intensity of a neuron is related to its number of signal emitted per unit time. The larger number of emision, the higher intensity.
    \item Small local patch of input neurons are aggregated into a vector, similar as the pooling in CNN but without max or average operation. 
    A patch of adjacent neurons are regarded as a signal source, emitting a number of identical vector signals at one timestamp. 
  \end{enumerate}
  Hence the three numbers in RGB channel are emision frequencies. The Blue channel in fig\eqref{fig:setup} shows the difference of signal frequency between dim grey and cornflower blue. 
  
  If the aggregated patch has size 3x3x3, then there would be $2^{27}$ states for the binary-input signals. 
  Choose the most confident state to be the target signal, corresponds to the fully-excited state. The reason that target signal having same shape as input signal is that similarity between them can be easily calculated. 
  The vector target is just the dense version of the usually used probability 1 scalar target. 
  
  Attention mechanism could occure between source signals and target signal. Target signal prefers to interact with similar signals above a threashold. 
  More discussion about attention interaction will be made when ising model is introduced later. 
  The green channel of fig\eqref{fig:setup} shows three sources at different physical locations; 
  the ratio between -1 and +1 element values in the signal vector would be different depending on the geometric shape of the plane. 
  Obviously the empty all-off signal would have the least similarity with target. 
  
  Unlike the routine back propagation learning where target is just one point in phase space, and doesn't have a physical location, 
  it is reasonable to allocate a physical coordinate for each class of target. 
  Because every source has a physical location which is different from the targets' locations, there could exist multiple routes between one pair of source and target. Signals must travel from sources to the target in order to proceed further calculations. 
  The red channel of fig\eqref{fig:setup} shows two sources, one on head and the other on wing of the plane. Signals emitted from them and received by the target signal with a specified location, could follow different routes. 
  Distances of routes as well as the travel speeds may be different; the resultant would be that the travelling time along different routes for any source-target pair are different. 
  Because two signals from the two sources are physically separated and are both excited by a same input picture, they must experience the same travelling time in order for them to arrive the target simultaneously. 
  Once the two signals arrived they can interact with the target signal to calculate energies either adding up or cancelling out each other, depending on their similarities with the target.  
  This is comparable with the light wave interference in optics, where photons arriving a location at one time would interfere to each other. The interference could be constructive or destructive, 
  depending on the phase angle similarity of the two photons. We could find a point on the interference pattern where constructive interference is the strongest. In our model this strongest interference point is desired to be the correct target's location after training is completed. 

  Trainig involves some trainable parameters, those parameters are the probabilities over different routes connecting one pair of of source-target. Therefore the targets representing different classes are not competing one probability; the competition is among the the routes. Any source must decide the probabilities of its travelling time towards any target. 
  Due to the existence of maximum speed limit, there could be travelling durations too short to make the trip possible. 

\subsection{Self attention as ising model}
Attention weights are higher between two similar values, indicating they are neighbours in phase space. A signal can be treated as a magnetic dipole in phase space, having similaries with other signals $s_{ij}=\boldsymbol{q}_i \cdot \boldsymbol{q}_j \in [-1, +1]$. 
Whenever the cosine-similarity in phase space exceeds a threshold, and also the two signals both exist at same time, attention interaction can occure. Interacting dipoles tend to point toward a same direction to lower the system's potential energy. 
$$
  \boldsymbol{q}_i \rightarrow mean(\boldsymbol{q}_i+\sum\limits_{ \begin{split} j \in \{sim(i,j) \\ 
  >threshold\} \end{split}} \boldsymbol{q}_j)
$$
can be considered as an evenly distributted attention weight being applied. 

Potential energy of an ising model is usually written as 

\begin{equation}
E=-\sum\limits_{i,j} \mathcal{J} \mathcal{\eta}(i,j) \boldsymbol{q}_i \cdot \boldsymbol{q}_j - \sum\limits_{i} \boldsymbol{q}_i \cdot \boldsymbol{B}_{ext}
\label{energy-ising}
\end{equation}
$\mathcal{\eta}(i,j)$ is 1 if $\boldsymbol{q}_i$ and $\boldsymbol{q}_j$ are similar else 0; $\mathcal{J}$ is the coupling coefficient, can be treated as constant=1. 
Ignoring the guest-guest interaction $\sum\limits_{i,j}$ term, treating target signal as the external field $\boldsymbol{B}_{ext}$, and only consider 
the interaction between dipoles and the target, is a simplified ising model. 

Applying attention mechanism between target and input signals, it is convenient to split signals into two groups using a similarity threshold:  
the attended group as similar friends, and others as the non-attending second group. Target is allowed to adjust himself under the influences of arriving friendly signals. 
Then the whole system's energy is calculated using this attended target.

Transformation of signal can occur any time between the signal's emision and its arrival time in the ising model event. Such transformations are ignored temporarily; in fact mean field theory 
can be used to explain the origin of non-linear transformations. This will be discussed in Further Work section, and now 
for simplicity signals are kept unchanged until they attend with the target.


  \subsection{Worldlines and travelling signals}
\begin{figure}[bth]
  \myfloatalign
  \subfloat[signal worldlines in GPT model]
  {\includegraphics[width=.45\linewidth]{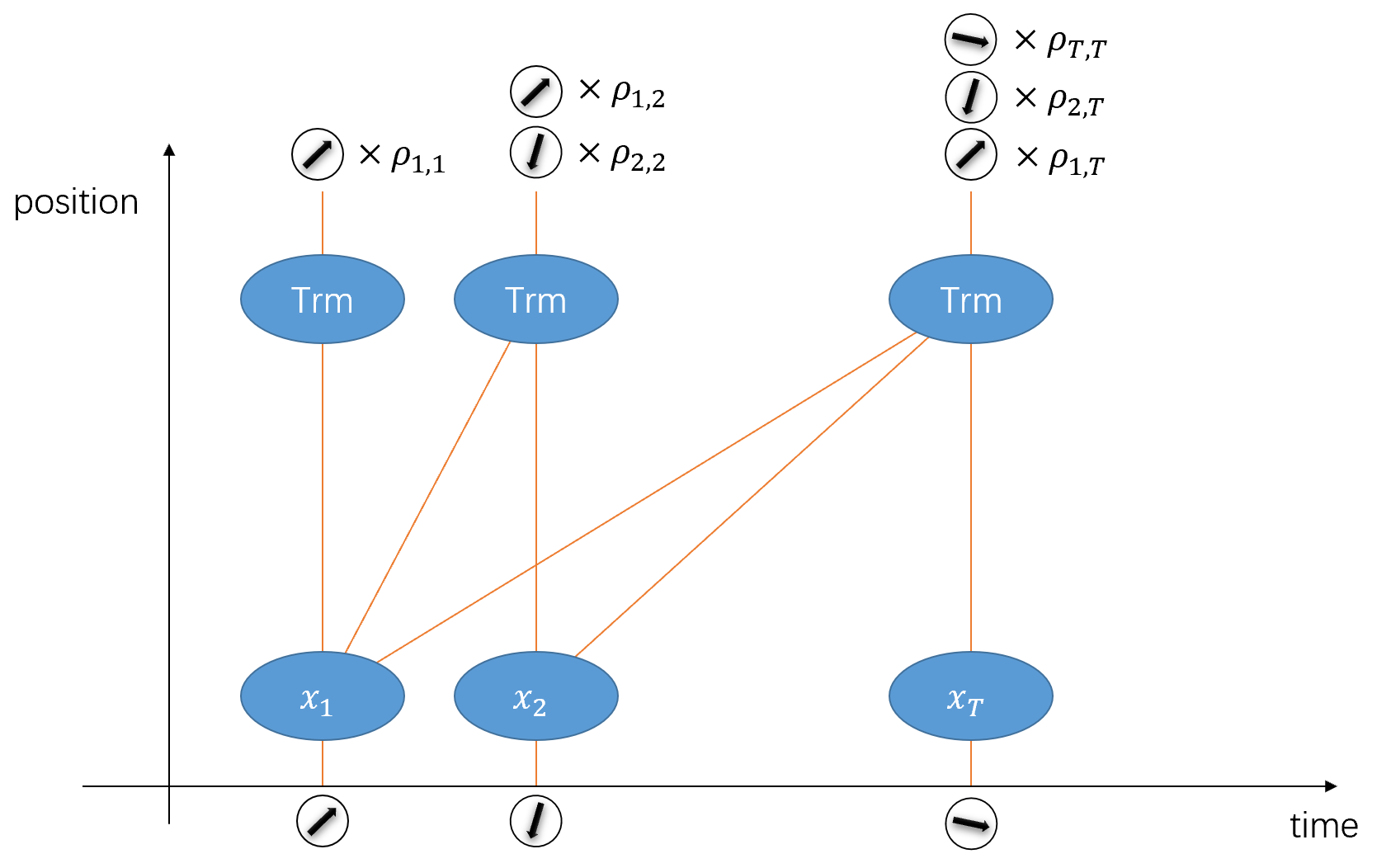}}
  \subfloat[external target field attend to similar signals in phase space]
  {\includegraphics[width=.45\linewidth]{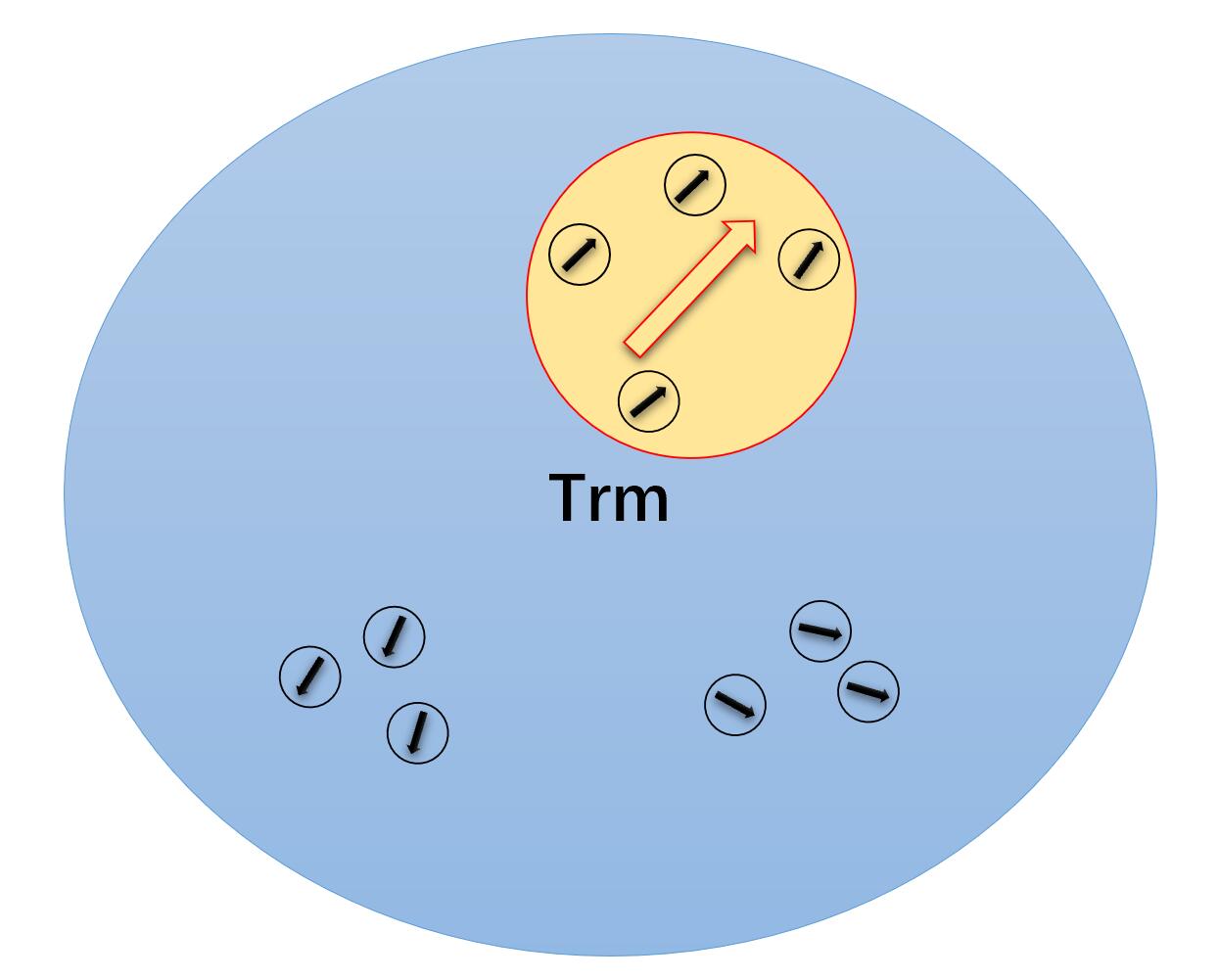}}
  \caption{gpt worldlines and interaction event} \label{fig:worldlines-gpt}
\end{figure}

  If the sources are dealing with time varying inputs, $\boldsymbol{x}(t)$ is like a RNN input in NLP tasks, then the travelling time becomes crutial to the self-attention mechanism. 
  Take one GPT layer, self-attention transformer receives input signals following different worldline trajectories. Fig\ref{fig:worldlines-gpt}(a) shows $\boldsymbol{x}_1$ signal need to travel at a lower speed compare with $\boldsymbol{x}_2$ speed, if both of them need to arrive transformer at time T. 
  Signal travel speed is reflected in the gradient of trajectories in space-time diagram. 
  Signals travelling from a source emitter to a target receiver, both have a fixed spatial location; the signal can choose multiple routes for this source-target combination, 
  each path may have a different length or a different conduction speed. Therefore route probabilities across different travelling time can be defined in the same way as in 2.1. 
  The assumption that conduction speeds for the routes are different, was inspired by the action potential in neuroscience, whose speed varies from 1m/s to over 100m/s \cite{hursh1939conduction}.  
  Noticing the route probability is not the same thing as self-attention weight. Self-attention weights are calculated within transformer as receiver at time T, whereas route probability is formed at emision time in the source input neuron position. 
  $\sum\limits_{i} a_{i,T}=1$ is the normalization of self-attention weights, whereas $\sum\limits_{t} \frac{\rho_{1,t}} {N} = 1$ is the normalization of route probabilities, N is the total number of signals emitted per unit time if neuron $x_1$ is activated. 

  In a self-interaction event, dipole signals suppose to interact with each other via a 'magnetic force'. Such a force is delivered by another type of signal which is a scalar, 
  acting in a short range, with speed much faster than the action potential. This high speed signal corresponds to the electrical signal in synapses, which is in charge of coordinating synchronized neuron excitation \cite{bennett2004electrical}. 
  Magnetic force is considered as instantaneous, therefore we assume the self-attention interaction is completed within a very short time that is negligible compare with the source-target travelling time.

  \subsection{Goal as function convolution}
  Method of calculating similarity between two signals is dot product.  
  Query signal is the external magnetic force experienced at target(destination) location, Keys are signals arriving target from different sources. A route is defined as a (source, target, travel\_duration) triplet. 
  Now we implement the same idea of attention weight update on emision route probabilities. If the arriving signals are similar to the target, then increase the corresponding route probability. The consequence would be if we randomly pick two signals arriving at same time in any self-attention event, 
  if they are similar, then source neurons will send more signals along the two paths with correct travelling duration, so that more signals would follow these paths and interact in the same 'meeting' event. 
  If query vector is always present and acting as a host, he can welcoming similar guests to join his meeting and expelling unwanted dissimilar guests since accepting those guests will cost more energy. 
  Define this process as \textbf{magnetic induction mutual attention}. After route probability update is completed, signals arriving an attention interaction 'meeting' will form a \textbf{constructive interference}. 
  Meaning the signals with direction closer to target, will superimpose on each other, therefore they would add up to reinforce each other instead of cancelling each other out. 
  
  If signals do not disappear immediately after his arriving 'meeting' event, it may interact with signals arrive in the future. 
  Our assumption is that signals must arrive simultaneously and are not allowed to survive after 1 timestamp, because old meeting event will be dismissed to allow the next meeting to start. 
  The reality may not be so strict. Two signals do not have to arrive simultaneously to enable self-attention interaction; as long as their worldline routes converge into a vacinity range in space-time diagram, 
  self-attention interaction can take place. This situation is ignored for simplicity reason. 
  
  Host signal is assumed to be present all the time. 
  Consider all the guest signals and their arrival time distribution, $\boldsymbol{f}(t) = \int\limits_{\tau = 0}^{t} \rho_{\tau,t} \boldsymbol{x}(\tau) d\tau$ is the resultant signal received by host at time t. $\rho_{\tau,t}$ is the emision number from source neuron at time $\tau$ and arrives attention interaction event held at time $t$. 
  Our goal is to maximize the peak of similarity between target signal and the received signal $J=\sup_{t \in [0, T]}\boldsymbol{f}(t) \cdot \boldsymbol{g}(t)$. This is the maximum instantaneous energy of the ising model. If the peak is high enough then hopefully the host will be able to emit signals to the next layer, or maybe it would trigger an avalanche of phase change in the ising model. For simplicity, assume the target host signal is
  $$
    \boldsymbol{g}(t) =\left\{
    \begin{array}{rcl}
    \boldsymbol{1} &  & 0 \le t \le T\\
    \boldsymbol{0} &  & \text{otherwise}
    \end{array}
    \right.
  $$
  Host signal is set to be a top-hat function with constant vector same size as signals $\boldsymbol{1}=(1, 1, ..., 1)^T$ during the interaction time period $[0, T]$. 
  Therefore the goal becomes a convolution function, which describes the strength of influence caused by impulse of arriving signals. 
  This is also the peak of energy-over-time function in our simplified ising model.
  $$J=\sup_{t \in [0,T]} \int\limits_{\tau=0}^{t} \rho_{\tau,t} \boldsymbol{x}(\tau) \cdot \boldsymbol{1} d\tau$$

  This goal applies to a single learner which can only response to one class. In order to solve a multi-class classification problem, we will need multiple learners.
  
  A modified mutual-attention mechanism would add a little time-dependent nudge to the target's direction. 
  This will affect the calculated energies of the meeting events.

\begin{figure}
  {\includegraphics[width=1.0\linewidth]{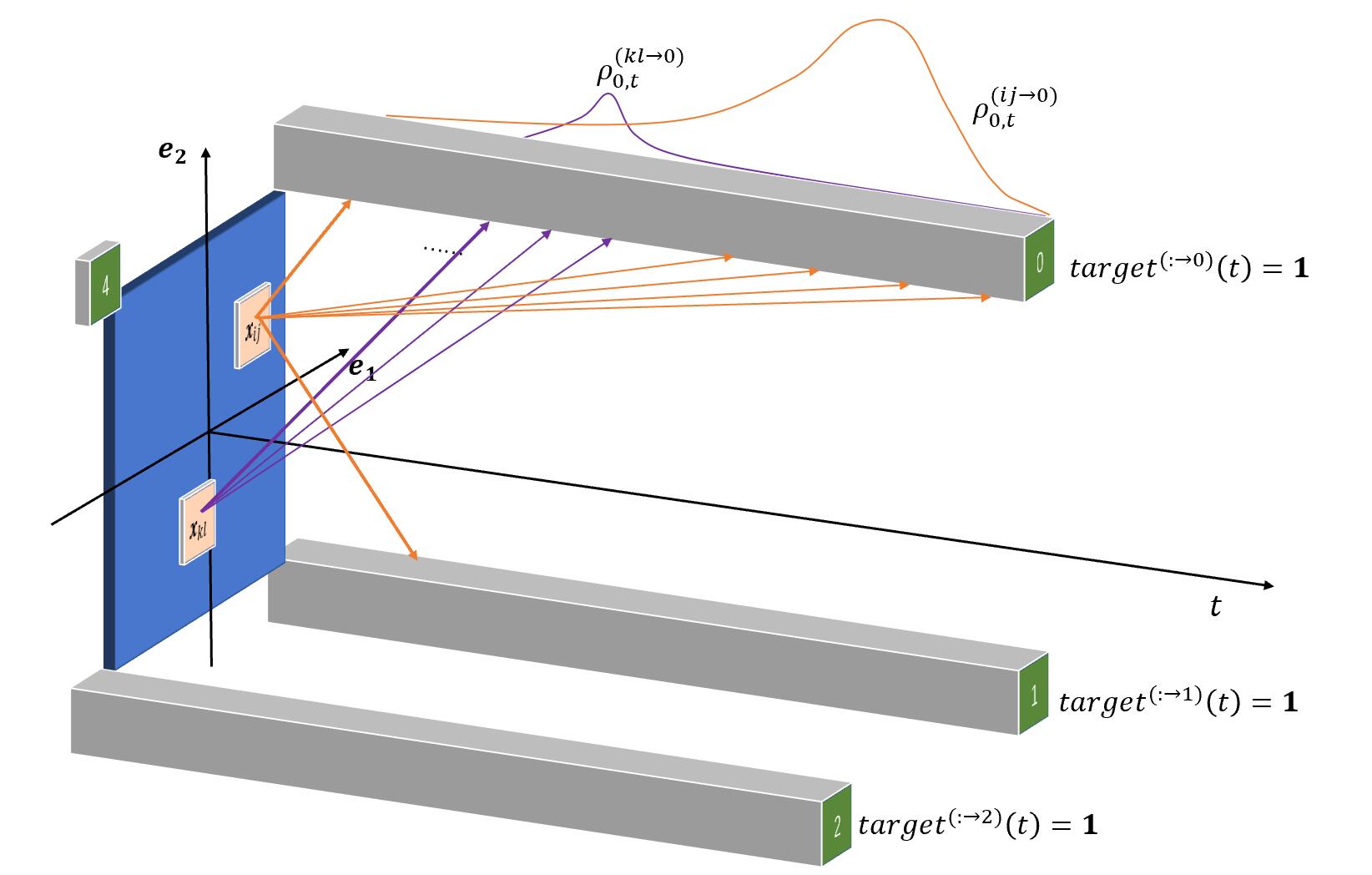}}
  \caption[4-class worldlines]{4-class worldlines}\label{fig:mnist-spacetime}
\end{figure}


\section{Method}

As shown in fig\ref{fig:setup}, the setup generates and propagates signals emitted from a 3D colored statue. The majority of datasets used in CV is not suitable 
because the 2D picture could not provide the correct spatial coordinates of the source patches relative to the geometric center as the origin. 
3D statue need to be constructed from a 2D picture before cutting into small patches of signal sources.  

MNIST dataset was chosen because:
\begin{enumerate}
  \item The structure is naturally 2D therefore statue 3D reconstruction is not necessary. 3x3 source patch locations and their constituent 9 binary values would not be obstructed from sight. 
  \item Orientations of the characters are roughly the same; rotations are already corrected before fed into the classifier.  
  \item Pixel values are roughly binary, this fits with our postulate about neuron excitation. 
  \item Characters are unicolor, only one channel is required. This keeps the interference and the magnetic induction procedure meanwhile simplifies the model. 
\end{enumerate}

4 classes were chosen out of 10 from MNIST dataset, $y \in \{0, 1, 2, 4 \}$. Spatial dimention was 2 in order to input features of a 2D picture, plus one temporal dimention to form our space-time diagram.
Preprocessing the picture required the picture to be truncated to size [27,27], cut into non-overlapping 3x3 patches, then reshaped to [W,H,9], where W=9 and H=9. Size of signal vector was 9+1, and there were WxH=81 different sources of signals. 
The extra one size of signal was used to represent the resting state, in case if all the 9 pixels are all 0, the normalized signal could still have modulus length equals to 1. $\boldsymbol{x}_{ij} \cdot \boldsymbol{x}_{ij} = 1 \forall i \in [1, 9], j \in [1, 9]$, $\boldsymbol{x}_{ij} \in \mathbb{R}^{10}$.
4 target classes were placed just outside the spatial corners of the picture as shown in Fig\ref{fig:mnist-spacetime}. 
Target signal also had its size increased by 1. 
$$target^{(:\rightarrow y)}(t) = \boldsymbol{g}_y = \begin{pmatrix} 0 \\ \frac{1}{3}\boldsymbol{1} \in \mathbb{R}^{9} \end{pmatrix} \forall t \in [0, T]$$
All signals were zero-centralized and then normalized, having modulus length equals 1. Noticing that anywhere else in this paper whenever a $\boldsymbol{1}$ appears, it really represents $\boldsymbol{g}_y$, the unattended normalized constant target signal, physically interpreted as the magnetic external field.

Manhattan distance was used to calculate spatial distances, signals' maximum speed was set to be 1. Every souce ij had its own route probability matrix. Time was truncated at T=24, enough for any signal to reach its target destination (maximum distance is 18), 
plus 6 timestamps to receive any slowers signals.  
Each route probability matrix had shape [4, 25], covering the 4 corner classes with any travel duration $\delta = t-\tau \in [0, 24]$. There were routes not possible of travelling because of the speed limit, and also there were routes detoured so signal would arrive later than expected. 
No noises nor transformations were applied to the travelling signals. 

Each training step, only one input picture was fed into the network, then system was allowed to evolve 24 timestamps, apply attention mechanism at each timestamp interaction event, meanwhile accumulating the route probability modification on each travelling duration. 
After 24 timestamps the accumulated modification was applied to the route probabilities, followed by a renormalizing process to preserve the probabilities. 

Mutual attention was applied between target host as query and input signals as keys to obtain the attended target:
$$
\boldsymbol{h}_{y}(t)=\sum\limits_{m(t)} \frac {s_{ym}} {\sum\limits_{m'(t)} s_{ym'}} \boldsymbol{v}_{m}
$$
where 
\begin{equation}
s_{ym} = \left \{ 
\begin{array}{rcl}
1 & & \text{if } \boldsymbol{q_y}\cdot\boldsymbol{k_m} \ge 0.7 \\
0 & & \text{elsewise}
\end{array}
\right .
\label{threshold-attn}
\end{equation}
This is just averaging the query with all keys having similarity greater than a threshold, under the constraint that signal could arrive the target location in time. 
$\boldsymbol{k}_m =\boldsymbol{v}_m = \boldsymbol{x}_{ij}(\tau)$ with source id $m(t):=9i+j \mid dist(ij\rightarrow y) \le t-\tau$; and the query is the target. $\boldsymbol{q}_y=\boldsymbol{g}_y = \boldsymbol{1}$.

After attention, target with all similar neighbours would have similarity equals 1, and the remaining signals would interact with the attended host target. Mutual interaction energy can be calculated on each arriving time t, and our goal was defined as: 
\begin{equation}
\begin{split}
J_y=\sup_{t \in [0,T]} \int\limits_{\tau=0}^{t} \sum\limits_{ij \mid s_{ym}=0}\rho_{\tau,t}^{(ij \rightarrow y)} \boldsymbol{x}_{ij}(\tau) \cdot \boldsymbol{h}_y(t) + \\
 \sum\limits_{ij \mid s_{ym}=1} \rho_{\tau,t}^{(ij \rightarrow y)} 1 d\tau
\end{split}
 \label{loss_eq}
\end{equation}

The goal was not maximised using gradients and back propagation. Instead we want to deliver all signals emitted from any source to the correct path, i.e. 
$\rho_{\tau,t}^{(ij \rightarrow y)} = N p_{\delta}^{(ij \rightarrow y)}$, The duration of travelling is $\delta=t-\tau$, N is the total number of signals emitted per unit time at current train step from source neuron ij towards target y, which is the frequency of the signal, and can be treated as constant in this experiment. 
As route probability $p_{\delta^*}^{(ij \rightarrow y)} \rightarrow 1$, with a suitable choice of travel duration $\delta^*$ on each source-destination combination (i,j,y), a signal can interact with as many 'similar friends' as possible, hence after the attention process, 
$$J_y \rightarrow \sum\limits_{ij \mid s_{ym}=1} \int\limits_{\tau=\tau_{emit}}^{t^*(i,j,y)=\tau_{emit}+\delta^*} N d\tau$$
because the first dot product term in Eq\ref{loss_eq} goes to 0 due to conservation of probability.

This implies if our goal is to minimise the infimum of negative potential energy (as an alternative way of saying maximising $J$), all \textbf{similar} signals chosen by host shall arrive at $t^*(y)=t^*(i,j,y) \forall i,j$ simultaneously, forming a sharpe spike on graph showing received signal number distribution over time, producing a low entropy state. 
Due to the presence of host signal $\boldsymbol{g}_y(t)$, generalized to be time dependent, $t^*(y)$ is expected to appear around the peak of $\boldsymbol{g}_y(t)$'s magnitude. 

$$
\rho^{:\rightarrow y}(\tau, t) = \sum\limits_{ij} \rho_{\tau, t}^{ij \rightarrow y}
$$
In the MNIST experiment signals were emitted only at $\tau=0$, therefore for target y, the similar signals' entropy is 
$$H_y=-\int\limits_{t=0}^T q(t) log q(t) dt$$ 
where $q(t) := \frac{\rho^{+\rightarrow y}(t)} {\int\limits_{t=0}^T \rho^{+\rightarrow y}(t) dt}$ and the superscript symbol '+' means all the similar signal sources. 
This entropy is expected to decrease in order for the similar signals to induce a larger instantaneous energy. 

However under the assumption that target external field is time-independent rather than a spiky impulse, the updating rule we were using was optimizing the energy during all time, 
not trying to concentrate the receiving peak, so there was a slight mismatch between our goal and the weight updating rule. Despite this difference the model could still learn, more detailed analysis will be done in the Experiment section. 

Route probability update method was simple, adding a modification term and then re-normalize over travel duration $\delta = t-\tau$: 
\begin{equation}
  p_{\delta}^{ij\rightarrow y}(new) = \frac{ p_{\delta}^{ij\rightarrow y} + \Delta w_{\delta}^{ij\rightarrow y}} {\int\limits_{\delta} p_{\delta}^{ij\rightarrow y} + \Delta w_{\delta}^{ij\rightarrow y} d\delta }
  \label{eq:update-rule}
\end{equation}
$$
  \Delta w_{\delta}^{ij\rightarrow y} = \left \{
  \begin{array}{rcl}
    \eta_{+} p_{\delta}^{ij\rightarrow y} & & \text{if } \boldsymbol{x}_{ij} \cdot \boldsymbol{g}_y \ge 0.7 \\
    \eta_{-} p_{\delta}^{ij\rightarrow y} & & \text{if } 0.7 > \boldsymbol{x}_{ij} \cdot \boldsymbol{g}_y > -0.7 \\
    \eta_{--} p_{\delta}^{ij\rightarrow y} & & \text{if } \boldsymbol{x}_{ij} \cdot \boldsymbol{g}_y \le -0.7
  \end{array} \right .
$$
where $\eta_{+}$, $\eta_{-}$, $\eta_{--}$ are learning rates for different similarities between arriving signals and target. 
Learning rates can vary for different number of training samples. For C=4, K=5 few-shot setup, $\eta_{+}=1.0$, $\eta_{-}=-0.5$, $\eta_{--}=-0.8$ were chosen. 
0.7 was a threshold defining the term 'similar'.

After the model was trained for 1 epoch, inference process was to calculate $J_y$ according to Eq\ref{loss_eq} $\forall y \in \{\text{label set}\}$ as if they were activated. The prediction is $y_{pred} = \mathop{\arg\max}_{y}  J_y$ 



\section{Experiment}

  \subsection{MNIST classification}

  Few training examples were randomly selected from MNIST, forming 3 train sets, each contain 1, 5, 10 pictures per class respectively. 
  One picture was fed into the network for each train step. Sequential train received inputs from different classes in sequential order; mixed train shuffled the input order. 
  For instance, with [a,b] being the two classes, SeqTrn(C=2,K=3)=[a,a,a,b,b,b], whereas MixTrn(C=2,K=3)=[a,b,b,a,b,a]. 
  The training example sequence was repeated for the number of epoches. 

  A fully connected neuron network with self-attention appled on input layer, then through a fully connected trainable weights of shape [4, 729], 
  predicting a softmax probability, using cross-entropy loss and Adam optimizer to back propagate the gradients. This one-layer model is the usual way of training, 
  and it was the compare benchmark for our proposed induced interference training.

  Test was evaluated on MNIST test set only using the 4 training digits [0,1,2,4]. 

  Our dynamic induced interference training outperformed the traditional back propagation model in few-shot scenario, 
  and it did not suffer from the catastrophic forgetting in sequential training. Result is in Table\ref{tab:ck-results}. In fact sequential learning can sometimes be more hepful. 
  For the one-shot situation, mixed training made no difference with sequential training, but the accuracies were largelly influenced by the quality of the one chosen picture. 
  Therefore the one-shot experiments were repeated for several times and then take the average accuracy. 
  Increasing K and epoch number can eventually let the BP network to achieve accuracies up to 94\% or higher, but clearly human learning doesn't require that much data and train steps. 
  
  Catastrophc forgetting is another important effect need to be addressed. Classical fully connect matrix transformed the common manifold where all input signals were projected on. 
  Learning one class would affect the shape of manifold and hence the relative distances between current input vector forwarded onto the hidden layer and the anchor vectors ($\boldsymbol{W}[y,:] \forall y \neq target$) of other classes, 
  therefore changing the output similarity with other classes. 

  In our ising model based forward propagate mechanism, route probabilities towards different target learners were normalized separately, 
  updating the probabilities towards $y=0$ doesn't affect the $y=1$ routes, therefore forgetting was not obvious, unless most of the two different class's input sources cannot distinguish the two target's space-time coordinates. 
  
  If the similar signals cannot arrive simultaneously, then the proportion of dissimilar signals in an attention event will increase. 
  This will cause the expected similarity between any two signals to decrease. This effect had been observed in the experiment; diagonal along Table\ref{tab:similar} 
  showed the estimated similarity metric after training has finished. 
  The supervising external target signal field had been withdrawn for this evaluation, so the second term in \eqref{energy-ising} was discarded, and the evaluation was interested on the first term. 
  In a learner's receiver view, probability distribution of arriving signal count over all sources can be calculated. This probability distribution became stable as more signals 
  were arriving towards the end of time T. 18 timestamps was the lower bound of travelling time for the furthest separated source-destination pair. 
  Left with 6 timestamps corresponding to 6 snapshots of complete-information interaction events. 
  Therefore each snapshot the expected similarity can be calculated as:
  $$
  \mathbb{E}_{q_l, q_m} S(t) = \sum\limits_{lm} q_l q_m \boldsymbol{x}_l \cdot \boldsymbol{x}_m
  $$
  where $l \in \{1...81\}$ loopping through source ids, and probability of receiving a signal from the source is
  \begin{equation}
  \begin{split}
  q_{l}(t) = \frac {\int\limits_{\tau}^t\rho_{\tau,t}^{i,j\rightarrow y} d\tau} {\sum\limits_{l} \int\limits_{\tau}^t \rho_{\tau,t}^{i,j\rightarrow y} d\tau} \\ 
  \forall l=9i+j
  \end{split}
  \end{equation}

  Then the expected similarity was averaged over $t \in [19,24]$ for a number of input examples selected from a desired label. 
  4 learners all had a largest expected similarity for their own label. This was a solid evidence that the inducing external field's precense 
  during training process was helping the input signals to show greater expected similarity only in the correct learner. 
  This was the same analogy as magnitizing a neutral iron block using a magnant. Take away the magnant, the iron block still exhibit magnetic field in the same direction as the magnant.

  Another interesting phenomenon was observed when analysing the expected signal arrival time shown by Table\ref{tab:arrival-time}. 
  We know the sources were sending signals with different travelling durations according to a distribution. A distribution was trained for each 
  target destination, so there were 4 trained signal sending policies, making up the route probability array. What would be the consequence of using a 
  wrong policy on any target? Obviously the optimizing goal would not be as high as if the correct policy is used, because constructive interference would be broken. 
  Apart from that, according to experiment, when white input was used, all sources send signals at same time, 
  the expected signal travel time using a wrong policy would be shorter than the travel time if the correct policy is used. 
  Although it was not always true when learners were re-trained, but most of the time this effect was observed. 
  It seemed to suggest that in human thought, correct answer usually takes longer time to be realized, and the first impression answer is likely to be wrong. 

\begin{table*}[]
  \centering
  \begin{tabular}{|l|l|l|l|l|}
  \hline
             & \begin{tabular}[c]{@{}l@{}}C=4,K=1\\ epoch=5\end{tabular} & \begin{tabular}[c]{@{}l@{}}C=4,K=5,\\ epoch=1\end{tabular} & \begin{tabular}[c]{@{}l@{}}C=4,K=10\\ epoch=1\end{tabular} & \begin{tabular}[c]{@{}l@{}}C=4,K=5\\ epoch=10\end{tabular} \\ \hline
  Our+SeqTrn & 66.63\%                                                   & 83.50\%                                                    & 88.65\%                                                    & 86.99\%                                                    \\ \hline
  Our+MixTrn & NA                                                        & 86.66\%                                                    & 82.48\%                                                    & 83.81\%                                                    \\ \hline
  BP+SeqTrn  & 38.41\%                                                   & 43.79\%                                                    & 35.04\%                                                    & 77.23\%                                                    \\ \hline
  BP+MixTrn  & NA                                                        & 53.28\%                                                    & 53.86\%                                                    & 79.99\%                                                    \\ \hline
  \end{tabular}
  \caption {few-shot experiment results}
  \label{tab:ck-results}
  \end{table*}

  \begin{minipage}{\textwidth}
    \begin{minipage}[t]{0.5\textwidth}
        \centering
        \begin{tabular}{|l|l|l|l|l|}
          \hline
          & y=0    & y=1    & y=2    & y=4    \\ \hline
 L0 & 0.2971 & 0.1876 & 0.0914 & 0.1559 \\ \hline
 L1 & 0.1551 & 0.4441 & 0.1507 & 0.1130 \\ \hline
 L2 & 0.0833 & 0.1168 & 0.1912 & 0.0534 \\ \hline
 L4 & 0.0951 & 0.1608 & 0.1258 & 0.2409 \\ \hline
\end{tabular}
\makeatletter\def\@captype{table}\makeatother\caption{expected signal similarity for last 6 timestamps}
\label{tab:similar}
    \end{minipage}
    \begin{minipage}[t]{0.5\textwidth}
    \centering
    
        \begin{tabular}{|l|l|l|l|l|}
          \hline
          & P0 & P1 & P2 & P4 \\ \hline
L0 & 15.99   & 9.49    & 10.17   & 7.24    \\ \hline
L1 & 18.05   & 18.49   & 13.63   & 12.75   \\ \hline
L2 & 17.24   & 14.28   & 17.97   & 17.51   \\ \hline
L4 & 9.72    & 10.31   & 9.89    & 15.43   \\ \hline
\end{tabular}
\makeatletter\def\@captype{table}\makeatother\caption{expected signal arrival time for sending policies}
\label{tab:arrival-time}  
    \end{minipage}
\end{minipage}



\begin{figure*}
  {\includegraphics[width=1.0\linewidth]{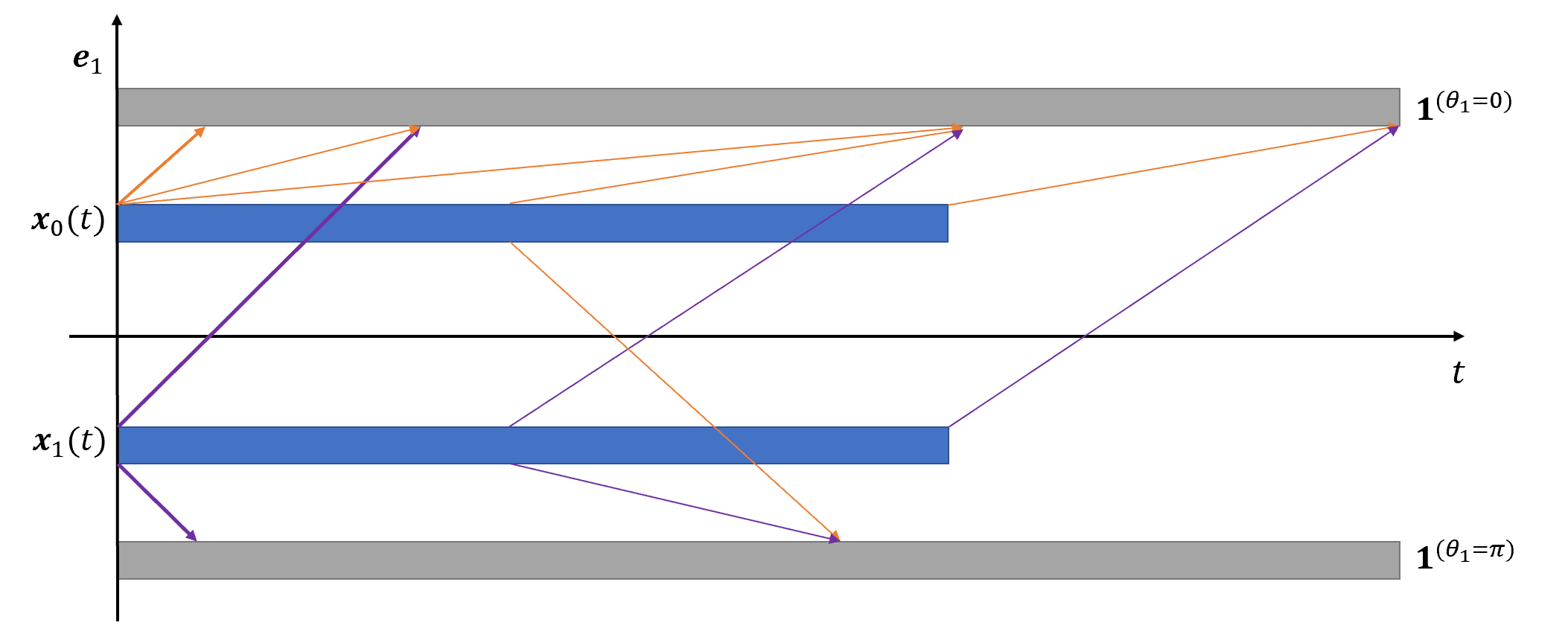}}
  \caption{double slit classification}\label{fig:double-slit}
\end{figure*}


  \subsection{Double slit experiment}
  With signal emision time spanning for a duration $L < T$, an experiment was firstly designed on a NLP task. 
  However the NLP task became trickey if word vecors were used as input. The main reason of failure is that word vectors are not made up by binary-like values. Therefore no fully excited state can be easily defined; i.e. 
  $\boldsymbol{wv} \cdot \boldsymbol{1} \neq similarity(word, target)$. Unlike the previous mnist experiment, where the 3x3 feature's activation level can be easily defined, 
  NLP word vectors has been transformed into a number of 'topics' equals to the number of vector size. Each 'topic' has a scalar 'magnitude' but its range cannot simply be stretched to $[-1, +1]$. 
  A viable representation is to use the crude one-hot representation as input, with size of dictionary equals the number of sources, each having a unique position. 
  This would produce a route probability tensor with shape $[N_{dict}, C, T]$, with dictionary size $N_{dict}$, number of classes C and travel duration with maximum value T. This is too large to be trained in practical; 
  for large picture/video inputs this can also be a serious problem for the training.
  Therefore some kind of aggregation+pooling method need to be used in order to reduce the number of signal travelling routes. Such a method will be proposed in the Further Work section. 

  In order to verify whether our induction model works for a time-varying signal source, A double slit experiment was designed to solve a 2-class classification problem. 
  The model was setup to identify whether two signal sources are in phase or out of phase, as shown in Fig \ref{fig:double-slit}. Two signal sources were rotating 2D unit vectors: 
  $$
  \boldsymbol{x}_{i}(t) = \left(
  \begin{array}{rcl}
  sin(\omega t + \theta_i + \epsilon)\\
  cos(\omega t + \theta_i + \epsilon)
  \end{array}
  \right)
  $$
  For the in-phase input, $\theta_0=\theta_1=0$; for the out-of-phase input, $\theta_0=0, \theta_1=\pi$. $\epsilon$ is a small random noise, $\omega$ is the coherent angular velocity of signal rotation. 
  Two targets were both unit vectors $\boldsymbol{y}=(0, 1)^T:=\boldsymbol{1}$. Similar as in the mnist experiment, one dummy dimension was added to represent the 'rest state'. Also the attended target $\boldsymbol{h}_y(t)$ was used to calculate the ising model energy.  

  The idea of the experiment was that if in-phase signals were emitted from input sources, then the route probabilities towards $\boldsymbol{1}^{(0)}$ were trained such that all signals arrived would form a constructive interference pattern at position of $\boldsymbol{1}^{(0)}$. 
  However this doesn't guarenteen a constructive interference in $\boldsymbol{1}^{(\pi)}$'s position with a different route probability policy when input signals remain in-phase. 
  Therefore both the energy averaged over time and the expectation of similarity would be the largest at $\boldsymbol{1}^{(0)}$'s position. 

  The opposite situation occured for the out-of-phase class. Therefore the two learners would be able to solve the classification problem. 

  Experiment showed expected results, model made predictions with 97\% accuracy.
  The code for both experiments (MNIST \& double slit) is released. \footnote{https://github.com/ttssrr423/InducedInterference}

~\\
~\\

\section{Further Work \& Discussion}
  \subsection{Multi-layer encoder}

  There are three things mentioned in earlier section which are not resolved. 
  \begin{enumerate}[]
    \item Dipole-dipole or guest-guest interaction among the arriving signals was ignored. This was an over-simplification.
    \item A method of aggregation+pooling need to be used to reduce the effective number of input sources. This reduces the number of routes need to be trained. 
    \item Non-linearity need to appear somewhere with a reasonable origination. 
    \end{enumerate}
  
  A hierachial attention mechanism is a proposed solution to the problems above. 
  Attention is normally used among vectors, however it can be generalized to be used among any ranked tensors. 
  Scalar attention $h_i=\sum\limits_{j} a_{ij} v_j$ is just perceptron model in another form, with scalar attention weight $a_{ij}(q_i, k_j)$ looked up from a trainable matrix. 
  A vector $\boldsymbol{v}_j^{(1)}$ with elements selected from set $H^{(0)}=\{h_1, h_2,...h_i,...\}$,  forms a subset $\boldsymbol{v}_{j'}^{(1)}=(h_1, h_2, ... h_D) \subset H^{(0)}$. Superscript (1) indicates the rank is 1, therefore a vector is aggregated from rank 0 scalars. 
  Now building the attention mechanism iteratively:

  $$
  \boldsymbol{h}_{i'}^{(1)} = \sum\limits_{j'} a^{(1)}(\boldsymbol{q}_{i'}^{(1)}, \boldsymbol{k}_{j'}^{(1)}) \boldsymbol{v}_{j'}^{(1)}
  $$
  $$
  =\sum\limits_{j'} a^{(1)}(\boldsymbol{q}_{i'}^{(1)}, \boldsymbol{k}_{j'}^{(1)}) \displaystyle\mathop{AGG}\limits_{j'\leftarrow \{i\}} \sum\limits_{j} a^{(0)}(q_i^{(0)}, k_j^{(0)}) v_j^{(0)}
  $$
  
  Superscript on attention weights are indicating the rank of tensor input pairs for the attention similarity function. 
  This equation is the general form of self-attention normally been used. 

  Repeat the aggregation-iteration a few more steps further, a rank R tensor can be treated as the output of the hierachial attention:
  $$
  \boldsymbol{h}_{i'}^{(R)} = \sum\limits_{j'} a^{(R)}(\boldsymbol{q}_{i'}^{(R)}, \boldsymbol{k}_{j'}^{(R)}) \displaystyle\mathop{AGG}\limits_{j'\leftarrow \{i\}} \boldsymbol{h}_i^{(R-1)}
  $$

  Dipole-dipole interactions are allowed to take place among the values in each ranked layer, using a threshold-based attention mechanism as Eq\ref{threshold-attn}. 
  Hence the previously ignored interaction is taken into consideration in a hierachial way. 

  The number of elements in $\boldsymbol{h}_{i'}^{(R)}$ is $D^R$ where D is the number of r-1 ranked tensor in a r ranked tensor. This exponentially increasing size will slow the computation until the forward propagate calculation becomes impossible. 

  The solution to the problem is to use mean-field approximation as the pooling method. In a high rank tensor this enables the dipole-dipole interaction to be modelled. 
  Approximation averages the attended values to produce a scalar, $\mu(\boldsymbol{h}_{i'}^{(R)}) = h_{i'}^{(R \rightarrow 0)}$. A useful property is that similarity between attended value $\boldsymbol{h}_{i'}^{(R)}$ and activation state $\boldsymbol{1}^{(R)}$ now can be calculated using a dot product operation, 
  because more activated elements/neurons will increase their parent tensor's mean field until it reaches the fully activated state $\mu(\boldsymbol{1^{(R)}})=1$. 
  The squashing of a tensor into a scalar used the idea of renormalization group, which allows similarities between tensor representations and target $\boldsymbol{1}$ to be approximately preserved across different scales. 
  Renormalization can be applied to a complex system not far from its critical state where self-similarity can be observed.

  Nested concepts with higher level of abstraction represented as a higher rank tensor (larger scale), is a much better way to encode words in tensors, compare with the word vector representation in NLP. 
  This not only fixes the problem of training NLP induced interference network (see 4.2 experiment section), but more importantly, gives rise to the non-linear response towards target signal, experienced as an external magnetic field. 
  Mean field operation reduces the number of input signal sources as well as the signal size. This enables the network to maintain fast forward propagate calculation speed even the attention rank level becomes high.

  Mean field solution of 2D square lattice ising model with energy described in Eq\eqref{energy-ising} is:

  \begin{equation}
    \mu = tanh\left( \beta(\bar{n}\mathcal{J} \mu + |B_{ext}|) \right)
    \label{MF2}
  \end{equation}
  With coupling coefficient treated as $\mathcal{J}=1$, and the mean field value of objective tensor being $\mu$. 
  $\bar{n}$ is the average number of attended neighbours in an attention 'meeting' event. $\beta=\frac{1} {k_BT}=const$ and $|B_{ext}|$ is the magnitude of magnetic external field on the induced field direction. 
  Target signal is no longer treated as time-independent, in fact host targets are not fundamentally different from guest signals. There can be multiple targets converge and interfere with each other to produce impulses with changing magnitude.
  During the estimation of mean field, signals are viewed as discretized binary valued vectors along the target-inducing direction in the viewpoints of higher ranked tensors. 
  This makes the 'opinion census' towards all elements under a tensor become faster.

\begin{figure*}[bth]
  {\includegraphics[width=1.0\linewidth]{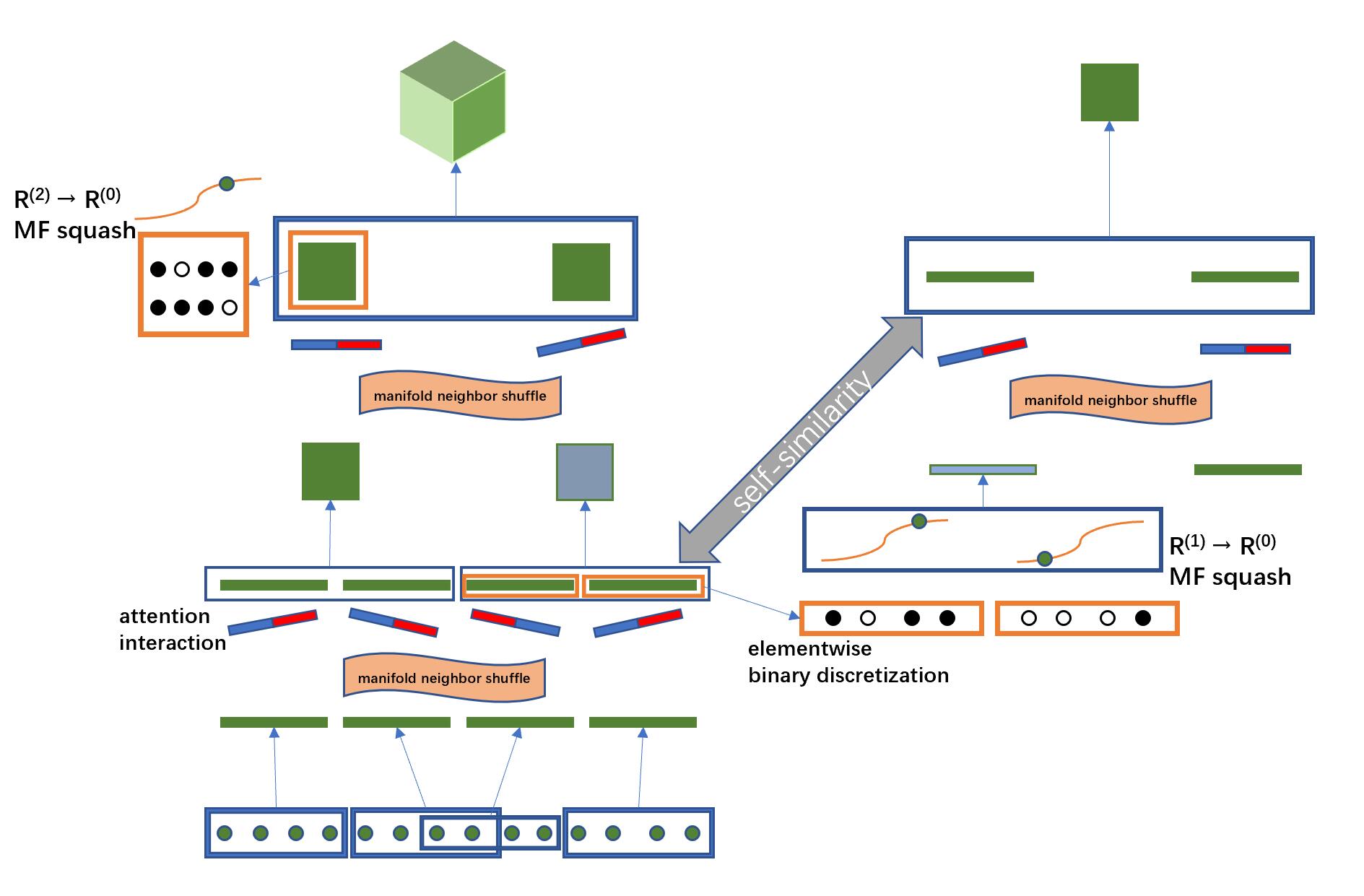}}
  \caption{Boxes with arrows show aggregations. Overlapping aggregation is allowed in rank 0. Left attention tower is unsquashed ranks, versus right tower is squashed on rank 1, using mean field approximation.}\label{fig:tower}
\end{figure*}

  The complete iterative estimation procedure is shown in Fig\eqref{fig:tower}. Manifold neighbour shuffle is applicable only if the kernel is no longer calculated as dot product; 
  if some parameterised kernel is used then the manifold would not be flat, hence similarity and the attention weights $a_{ij}(q_i, k_j)$ will also be influenced, causing the attended neighbours to be changed. 


\begin{figure*}[bth]
  \myfloatalign
  \subfloat[tanh-like function for $\beta \bar{n} J < 1$]
  {\includegraphics[width=.45\linewidth]{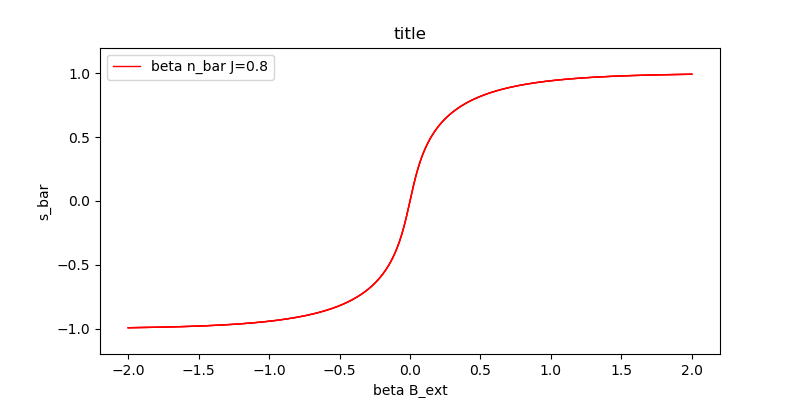}} \quad
  \subfloat[phase change for $\beta \bar{n} J > 1$]
  {\includegraphics[width=.45\linewidth]{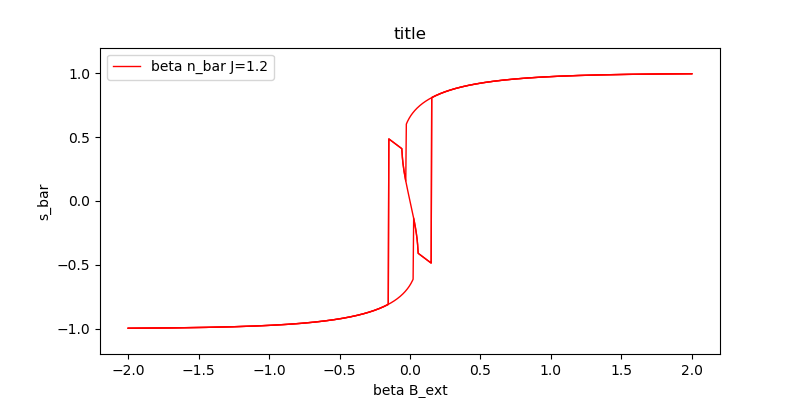}} \\

  \caption{non-linear activation function plotting $\mu$ against $\beta B_{ext}$}\label{fig:transCompare}
  \end{figure*}

  Plotting Eq\eqref{MF2} with $\beta |B_{ext}|$ on x-axis and mean-field $\mu$ on y-axis, there are two types of activation function obtained depending on whether the value of $\beta\bar{n} > 1.0$ or not. 
  Our experiment in (4.1) has proven the expected similarity between arriving signals will increase after training. 
  Therefore the number of similar signals exceeding threshold will also increase, causing the dipole-dipole interaction having more average number of similar neighbours, represented by $\bar{n}$. 

  There would be a point where the continuous activation function shown in Fig\ref{fig:transCompare}(a) becomes a discontinuous phase-transition allowed activation function shown in Fig\ref{fig:transCompare}(b) as $\bar{n}$ increases. 

  \subsection{Dissipative structure} 
  The most signifficant effect of phase transition is that small change in the external field can cause huge change in the induced mean field. 
  Tensors are nested in the hierachial structure and each tensor may exist in its critical state where phase change can be triggered with a small perturbation on the environment. 
  If there are a large proportion of tensors already in their critical state, then avalanche of phase transition would occure easily. 

  Energy dissipated into the surrounding environment in one switching cycle of mean field output, is proportional to the area enclosed by hysteresis in \ref{fig:transCompare}(b). 
  After the avalanche is stopped and the network reaches a new stable state, the entropy of the network is expected to be reduced. Dipole signals would interact in space-time diagram events with 
  more constructive interference and therefore having a lower entropy. As a result of second law of thermodynamics, the environment receives the dissipated energy and 
  the entropy of environment would increase to ensure the total entropy of universe $H_{universe}=H_{system}+H_{environment}$ to increase. 

  Sandpile model \cite{PhysRevLett.59.381} is a good analogy to model this avalanche of phase transitions. High entropy input signals with energy provided by signal sources would 
  forward propagate through different distanced routes to be fed into the hierachial attention tensor. Different abstraction level (on different ranks) of tensors would estimate 
  their own mean field output, to complete the pooling process and then sending approximated signals to new destinations. Meanwhile route probabilities are updated by simple rule 
  described in Eq\ref{eq:update-rule}. This update enables the mean field attentions to have a lower entropy most of the time; also it doesn't require information back propagated from top-layer, 
  but only requires the nearby external magnetic field, so updating can be done locally. Because the local target field can change with time, it is possible that 
  within an attention tensor, phase change can produce discontinuous output, triggering phase change in other tensors connected with it after the signal travelling time. 
  An avalanche can propagate within a much larger region of brain just like the sandpile model. Self-organization is expected to be observed after avalanche so that the network system can maintain itself by comsuming 'negative entropy' \cite{schrodinger1962life}.


~\\
~\\

\section{Conclusion}
  Self-attention mechanism has been proven successful in rescent year transformer-based encoders. By realizing the physical interpretation of self-attention being signal interactions, 
  ising model becomes a viable choice of modelling a learning process. Despite the forces governing the interaction remain unknown, its physical property is assumed to be similar as magnetic force. 
  Such force is assumed to be transmitted via a fast-speed signal, i.e. the electrical signal pass along neurons. Comparing with the slow travelling action potential signal speed which was used to model the interfering signals.
  The time cost for attention interaction to complete within a tensor would be much shorter than the slow signal travelling time from sources to targt tensors. 
  By fixing positions of signal sources and target region, then setting signal travel speed to have an upper bound, space-time condition where an ising model interaction event takes place can be identified. 
  This brings the idea of travelling routes and constructive interference defined same as in optics. Training process is trying to make constructive interference among input and target signals at a specified position. 
  Once route probabilities are trained, no matter whether target signal is still present or not, constructive interference would retain, because the input signals would still interfere with each other. 

  A constructive interference also suggests the potential energy of system is minimized, therefore by ignoring the dipole-dipole energy, but keeping the dipole-target energy, an simplified energy-based goal was proposed. 
  Probabilities of routes with different travelling time were updated for each source-destination pair, in the aim of optimizing our goal. 

  Two experiments were designed. One did a 4-class MNIST classification, showing good few-shot performance, and it was immune to catastrophic forgetting. 
  It also confirmed the induction by target signal was effective, as the expected dipole-dipole similarity was highest in the correct target's position even if the target signal was absense. 
  The second experiment did a in-phase/out-of-phase periodic wave pattern binary classification, demonstrated time varying signals can also be learned using this interference idea. 

  A failed NLP task attempt lead to the reconsideration of using word vectors. A nested aggregation, mean field pooling and attention interaction mechanism was proposed to bridge ising model and concept nesting. It brought the following features into our learner. 

  \begin{enumerate}
    \item It has a fractal structure, with self-similarity between different ranks and different scales.
    \item Mean field pooling lowers the number of updating routes towards the next destination if the tensor learner is relaying signals. This makes future calculations to be more efficient. 
    \item Mean field brings back the previously ignored dipole-dipole interaction. It allows the tensor learner to have a critical state. The constituent sub-ranked tensors can also have their own critical states. 
    \item Avalanche of phase transition can occure and propagate throughout the network system containing a large number of learners. Avalanche magnitude versus frequency graph may not obey the power law, because we know human brain is actively maintaing the brain's critical state \cite{brochini2016phase}. Therefore it is reasonable to assume that the stopping condition of training is to keep the tensors in network not too far from their critical states. 
    \item Non-linearity and discontinuity emerge as a consequence of phase transition. 
    \end{enumerate}
  
  All the evidences are pointing towards dissipative structure and edge of chaos. It would be interesting to study signal interference in such a complex system.

\bibliographystyle{unsrt}
\bibliography{reference}
\end{document}